# Magnetic and magnetocaloric properties of melt-extracted $Mn_{1.26}Fe_{0.60}P_{0.48}Si_{0.52}$ microwires

Lin Luo[1], Hongxian Shen[1,2], Sida Jiang[1], Ying Bao[1], Yongjiang Huang[1], Shu Guo[1], Ze Li[1], Nguyen Thi My Duc[2,3], Jianfei Sun[1,*], and Manh-Huong Phan[2,*]

[1] School of Materials Science and Engineering, Harbin Institute of Technology, Harbin 150001, China
[2] Department of Physics, University of South Florida, Tampa, Florida 33620, USA
[3] The University of Danang, University of Science and Education, 459 Ton Duc Thang, Lien Chieu, Danang, Vietnam

The polycrystalline $Mn_{1.26}Fe_{0.60}P_{0.48}Si_{0.52}$ microwires were successfully fabricated for the first time by the melt-extraction technique, and their magnetic and magnetocaloric properties were investigated systematically. The structural analysis shows that the microwires possess a hexagonal phase with $Fe_2P$ type, with a homogeneous composition distribution. Magnetometry measurements show that the microwires undergo a weak first-order magnetic phase transition (FOPT) at a temperature of ~142 K. The maximum magnetic entropy change ($\Delta S_M^{pk}$) of the microwires reaches ~4.64 J/kg·K for a field change of 5 T. These low-cost $Mn_{1.26}Fe_{0.60}P_{0.48}Si_{0.52}$ microwires are promising for active magnetic refrigeration in the liquid nitrogen temperature range.

* Corresponding authors: hitshenhongxian@163.com (H. X. Shen), phanm@usf.edu (M. H. Phan), jfsun@hit.edu.cn (J. F. Sun)

## 1. Introduction

Magnetic refrigeration, which is based on the magnetocaloric effect (MCE), has attracted growing interest due to its high efficiency and environmental friendliness [1-5]. This technology is considered a potential alternative to the conventional vapor-compression techniques. However, the development and the application of magnetic refrigerators depend on the performance of magnetic refrigerants, especially for high temperature regimes around room temperature [5]. Some alloys that possess the giant MCE near room temperature, such as $Gd_5(Ge, Si)_4$, $La(Fe, Si)_{13}$, Heusler alloys and $(Mn, Fe)_2(P, Si)$, have triggered a great deal of attention in advancing the magnetic cooling technology [6, 7]. However, their practical application is limited by several factors, such as finite natural resources, high cost of rare-earth (RE) metals, poor mechanical properties, and a complex assembly of irregular refrigerant parts [5]. Thus, there is a pressing need for developing new materials that have a combination of the giant MCE and low-cost without RE elements.

The $(Mn, Fe)_2(P, Si)$ alloys with giant MCEs are considered one of the most promising refrigerants, due to the low-cost and abundance of the raw materials, which are potentially beneficial for large-scale magnetic cooling application [8]. Over the past few years, many MCE studies have been concentrated on these materials. Dung *et al.* [9] reported that the Curie temperature ($T_C$) and the magnetic entropy change ($\Delta S_M$) reduced when the Mn/Fe radio decreased in $(Mn, Fe)_{1.95}P_{0.5}Si_{0.5}$ ($x \geqslant 1.1$) compounds. They also showed that the hysteresis could be tuned by varying the

Mn/Fe radio. Katagiri *et al.* [10] found that the $T_C$ increased linearly with increasing $x$ from 0.3 to 0.6 in MnFeP$_{1-x}$Si$_x$ compounds. The effects of doping and heat treatment on the MCE performance have been intensively investigated [11-16].

It is worth noticing that most of the (Mn, Fe)$_2$(P, Si) alloys were prepared by powder metallurgy and/or vacuum melting methods. However, these methods are usually time-consuming, up to several days, for achieving a homogeneous composition. On the other hand, the rapid solidification has attracted growing attention as it enables the fabrication of materials with excellent composition homogeneity and the reduction of the heat treatment time [5,17]. Ou *et al.* [18] prepared Mn$_{0.66}$Fe$_{1.29}$P$_{1-x}$Si$_x$ ribbons by the melt-spinning technique and quenched the ribbons into water after annealing for only 2 h. The melt-extraction technology applied to fabricate microwires possesses a larger solidification rate (up to $10^6$ K/s) [19]. Previous reports showed that the La(Fe,Si)$_{13}$ microwires fabricated by the melt-extraction technology only required a few minute heat treatment for achieving a homogenous element distribution, while their bulk and ribbon counterparts prepared by conventional vacuum melting and melt spinning required several days and a few hours for heat treatment, respectively [5, 20]. In addition, theoretical calculations have shown that the wire-shaped magnetic refrigerants with close packing and micro-size diameter yield a higher heat exchange efficiency with an enhanced surface area and a higher working frequency, as compared to their bulk and film counterparts [21, 22]. A large body of work on the MCE performance of melt-extracted Gd alloy microwires has been reported [19, 23-35]. It is therefore expected that the melt-extraction technology is most appropriate for fabricating (Mn, Fe)$_2$(P, Si) wires with good MCE performance.

In this work, we report on the successful fabrication of Mn$_{1.26}$Fe$_{0.60}$P$_{0.48}$Si$_{0.52}$ microwires using the melt-extraction method. The macroscopic feature, structural information, magnetic and magnetocaloric properties of the microwires were systematically investigated. Magnetocaloric studies show that the Mn$_{1.26}$Fe$_{0.60}$P$_{0.48}$Si$_{0.52}$ microwires are promising for active magnetic refrigeration in the liquid nitrogen temperature range.

## 2. Experimental

The master alloy with a nominal composition of Mn$_{1.26}$Fe$_{0.60}$P$_{0.48}$Si$_{0.52}$ was prepared by arc-melting a mixture of Mn (99.5%), red P (99.5%), FeP chunks (99%), and Si lumps (99.999%) in a Ti-gettered high-purity argon atmosphere. An excess 5% of Mn was added to compensate for the loss during melting. The ingot was re-melted at least six times with electromagnetic stirring to ensure chemical homogeneity, and then suction casted to a rod sample with diameter of ~8 mm in copper mold. Subsequently, we put the alloy rod into a self-designed precision melt-extraction equipment and re-melted it using electro-magnetic induction. The melt feeding speed was fixed at 20 μm/s. The copper wheel with a diameter of 300 mm was employed as melt spinning, and the line speed of the wheel rim was fixed at 30 m/s. When the melt contact with the rim edge of the high-speed spinning copper wheel, the

$Mn_{1.26}Fe_{0.60}P_{0.48}Si_{0.52}$ microwires were extracted and casted.

X-ray diffraction (XRD) was performed to characterize microstructure of the microwires. The microwires were cut into small pieces (~ 2 mm) and occupied inside a glass sample holder uniformly. The XRD patterns were collected from 10° to 80° using a D/max-rb with Cu-Kα radiation on an X'pert facility produced by Philips. The macro morphology was observed on a field emission scanning electron microscope (SEM-SIGMA-500-Zeiss), and the chemical composition was analyzed on energy dispersive spectrometer (EDS) in SEM. The magnetic measurements were carried out on a Quantum Design Physical Property Measurement System (PPMS-16 T) with a vibrating sample magnetometer (VSM) module. To avoid the thermal-history effect, the samples were precooled to 10 K under zero magnetic field before each magnetic measurement, and the isothermal magnetization measurements were performed using the so-called loop process [36].

## 3. Results and Discussion

The XRD pattern of the $Mn_{1.26}Fe_{0.60}P_{0.48}Si_{0.52}$ microwires is displayed in Fig. 1. The $Mn_{1.26}Fe_{0.60}P_{0.48}Si_{0.52}$ microwires show an obvious $Fe_2P$-type hexagonal structure (space group of *P¯62m*). This is consistent with that reported in previous studies [37, 38]. Meanwhile, small weak peaks indicated that the samples present a minor impurity phase of $(Mn, Fe)_5Si$ (space group of *P63/mcm*).

The length of the as-cast microwires was between 5 and 10 cm. However, the average diameter of the microwires was ~30 µm. The macroscopic feature of the casted microwires is shown in SEM images (see Figs. 2(a) and (b)). Groove defect existed in the area contacted with the copper wheel during the solidification process. It is noted that the grains in these microwires are fine, and their size is less than 5 μm. Three points in Fig. 2(a) were chosen for EDS tests, to determine the chemical composition. The result, as shown in Fig. 2(c), indicates that the composition of the microwire is $Mn_{1.26}Fe_{0.60}P_{0.48}Si_{0.52}$. From Fig. 2(b), the corresponding EDS mapping of Mn, Fe, P, and Si elements was performed, and the results are shown in Figs. 2(d)-(g). The distribution of chemical compositions in the as-cast microwires is homogeneous.

The temperature dependent magnetization (*M-T*) curves for the $Mn_{1.26}Fe_{0.60}P_{0.48}Si_{0.52}$ microwires were measured under a low magnetic field (0.1 T), as displayed in Fig. 3(a) and its inset. As one can see in this figure, the *M-T* curve exhibits a broad paramagnetic to ferromagnetic (PM-FM) phase transition around the Curie temperature $T_C$. The $T_C$, which is defined as the corresponding temperature of the peak value of d*M*/d*T*, is determined to be ~142 K (not shown here). However, the determination of $T_C$ using this method is inappropriate for materials exhibiting a very broad magnetic ordering phase transition [33]. According to the modified Bloch's law, the *M-T* curve at the low temperature region can be well-fitted by the following equation [39]:

$$M(T) = M(0) \times \left[1 - \left(\frac{T}{T_C}\right)^{\alpha}\right], \tag{1}$$

where $M(0)$ is the magnetization of the ground state at $T = 0$ K, $T_C$ is the Curie temperature, and α is the Bloch exponent. Fig. 3(a) also shows the modified Bloch's law fitting curve (the red curve) and its inset shows a magnified graph over a temperature range from 0 to 140 K. The $M(T)$ dependence obtained by fitting equation (1) decreases with increasing temperature. The parameters obtained from the modified Bloch's law fitting curve are $M(0) = 70.44 \pm 0.04$ (A m$^2$ kg$^{-1}$), $T_C = 174.40 \pm 0.25$ K, and $\alpha = 2.628 \pm 0.007$. The Bloch's law suggests that the decrease in the magnetization with increasing temperature is due to spin-wave excitations.

In the paramagnetic region, the susceptibility should follow the Currie-Weiss law,

$$\chi = C/(T - \theta_P) \quad , \tag{2}$$

where $C$ is the Curie constant, and $\theta_p$ is the paramagnetic Curie-Weiss temperature. The temperature dependence of the inverse susceptibility is shown in Fig. 3(b). It is noted that the susceptibility deviates from Curie-Weiss law up to ~260 K. This deviation, which was commonly observed in hexagonal Fe$_2$P-based compounds, could be attributed to the presence of short-range magnetic order [40]. The values of $C$ and $\theta_p$ were obtained by fitting the paramagnetic region to Currie-Weiss equation. $\theta_p$ is determined to be ~206 K, which is higher than $T_C$ ~142 K. This also demonstrates the presence of magnetic inhomogeneity around 60 K above $T_C$ [41, 42]. From the value of $C$, the effective moment per formula unit $\mu_{eff}$ can be determined by the following equation [41],

$$C = \frac{N}{3k_B}\mu_{eff}^2 \quad , \tag{3}$$

where $N$ is the number of Avogadro and $k_B$ is the Boltzmann's constant. Then, the calculated value of $C$ is 5.31 emu K mol$^{-1}$ and $\mu_{eff}$ is 6.52$\mu_B$. This high value of $\mu_{eff}$ can be attributed to the presence of short-range ferromagnetic clusters.

Fig. 4(a) presents the filled 2D contour plot of the temperature and applied magnetic field dependence of magnetization for the Mn$_{1.26}$Fe$_{0.60}$P$_{0.48}$Si$_{0.52}$ microwires to provide a greater clarity on the $M(T)$ broadening. From Fig. 4(a), the magnetization can be seen to show a significant change in the temperature region from 125 K to 170 K, where the $T_C$ has been calculated (~142 K) from the minimum of the d$M$/d$T$. This leads to an expectation that the maximum entropy change will occur in this temperature region. To highlight the phase transition region in the d$M$/d$T$ curves, the filled 2D contour plot of the temperature and applied magnetic field dependence of d$M$/d$T$ for the Mn$_{1.26}$Fe$_{0.60}$P$_{0.48}$Si$_{0.52}$ microwires is shown in Fig. 4(b). The purple color area in Fig. 4(b) clearly indicates that the lowest value of d$M$/d$T$ varies over a temperature range of 125 - 170 K.

In order to analyze the MCE performacne of the Mn$_{1.26}$Fe$_{0.60}$P$_{0.48}$Si$_{0.52}$ microwires, we measured a set of isothermal magnetization curves $M(\mu_0 H)$ at given temperatures ranging from 10 to 250 K in magnetic fields up to 5 T (not shown here). The magnetic entropy change (-$\Delta S_M$) can be determined from these isothermal magnetization curves via the Maxwell relation [43]:

$$\Delta S_M = \int_0^{H^{max}} \left(\frac{\partial M}{\partial T}\right)_H dH \quad . \tag{4}$$

The 3D colormap surface and the filled 2D contour plots of temperature and magnetic field change (up to 5 T) dependencies of -$\Delta S_M$ for the Mn$_{1.26}$Fe$_{0.60}$P$_{0.48}$Si$_{0.52}$ microwires are shown in Figs. 5(a) and (b), respectively. The value of the peak magnetic entropy change for a field change of 5 T is 4.64 J kg$^{-1}$ K$^{-1}$, which is lower than that of the MnFePSi alloys with a similar composition [9, 37]. It has been reported that the MnFePSi-type alloy series exhibit a gradual change from the first-order to second-order magnetic transition [44]. The lower values of MCE obtained for the Mn$_{1.26}$Fe$_{0.60}$P$_{0.48}$Si$_{0.52}$ microwires may suggest the presence of a weak first-order magnetic transition, which could be related to the presence of small grains in the microwires [45].

The cooling efficiency of the Mn$_{1.26}$Fe$_{0.60}$P$_{0.48}$Si$_{0.52}$ microwires has also been estimated. Typically, the three methods are generally used to characterize the cooling efficiency of magnetocaloric materials [46]:

$$RC1 = -\Delta S_M^{peak} \times \delta T_{FWHM} = -\Delta S_M^{peak}(T_2 - T_1) \tag{5}$$

$$RC2 = \int_{T_1}^{T_2} -\Delta S_M(T) dT \tag{6}$$

$$RC3 = (-\Delta S_M \times \delta T)_{max} \tag{7}$$

where the $\delta T_{FWHM}$ is the full width at half maximum of a -$\Delta S_M$ ($\mu_0 H$, $T$) curve, the $T_1$ and $T_2$ are the corresponding onset and offset temperatures of $\delta T_{FWHM}$. The applied magnetic field dependences of RC1, RC2, and RC3 are displayed in Fig. 6. The calculated values of RC1, RC2, and RC3 are ~441, 343, and 222 J kg$^{-1}$ for a field change of 5 T, respectively. These results show the good magnetocaloric response of the $Mn_{1.26}Fe_{0.60}P_{0.48}Si_{0.52}$ microwires.

To determine the type of the magnetic phase transition in the melt-extracted $Mn_{1.26}Fe_{0.60}P_{0.48}Si_{0.52}$ microwires, the Arrott plots derived from the isothermal magnetization curves were employed. According to Banerjee criterion, a magnetic transition is of second-order type if all the slopes of the Arrot plots are positive; otherwise, it is of first-order type [24, 47]. All Arrott plots for the $Mn_{1.26}Fe_{0.60}P_{0.48}Si_{0.52}$ microwires show a positive slope (Fig. 7(a)), indicating a typical character of second-order magnetic transition. This seems inconsistent with the MCE analysis that reveals the weak FOPT in the $Mn_{1.26}Fe_{0.60}P_{0.48}Si_{0.52}$ microwires.

To resolve this discrepancy, the universal $-\Delta S_M$ ($\mu_0 H$, $T$) curve method has been employed [48]. The construction of a phenomenological universal curve can be described as plotting $\Delta S / \Delta S_M^{peak}$ vs. $\theta$ curves. The temperature axis, $\theta$, is defined as:

$$\theta = \begin{cases} -(T - T_{peak})/(T_{r1} - T_{peak}) & T \leq T_{peak} \\ (T - T_{peak})/(T_{r2} - T_{peak}) & T > T_{peak} \end{cases}, \quad (8)$$

where $T_{r1}$ and $T_{r2}$ are the reference temperatures ($T_{r1} < T_C$ and $T_{r2} > T_C$), and they satisfy the relationship for each value of the applied field and any arbitrary value of $h$:

$$\Delta S_M(T_{r1})/\Delta S_M^{peak} = \Delta S_M(T_{r2})/\Delta S_M^{peak} = h, \quad (9)$$

In this work, $h = 0.8$ was selected to construct the universal curve. In this phenomenology approach, the order of the transition has a decisiveness effect only below $\theta = -1$. The dispersion is calculated by:

$$\text{dispersion} = 100 \times \frac{W(\theta=-5)}{\Delta S_M / \Delta S_M^{peak}(\theta=-5)}, \quad (10)$$

where $W$ is the width of the vertical spreading. The dispersion is more than 100% for compounds with first-order phase transition and less than 30% for compounds with second-order phase transition. The phenomenological universal curves are shown in Fig. 7(b). All the curves display an obvious breakdown at $\theta <-1$, and the calculated dispersion is 116%, more than 100%, indicating the presence of first-order magnetic phase transition. We thus speculate that the Banerjee criterion is not appropriate for analyzing the nature of the magnetic phase transition in the melt-extracted $Mn_{1.26}Fe_{0.60}P_{0.48}Si_{0.52}$ microwires. Nevertheless, the first-order transition in these microwires is rather weak, and the sample displays more characteristics of a second-order magnetic phase transition [49,50].

## 4. Conclusion

In summary, the melt-extracted $Mn_{1.26}Fe_{0.60}P_{0.48}Si_{0.52}$ microwires were successfully fabricated by the melt-extraction technique. The structural, microstructural, magnetic, and magnetocaloric properties were investigated. The $Mn_{1.26}Fe_{0.60}P_{0.48}Si_{0.52}$ microwires possess a $Fe_2P$-type hexagonal structure, with a minor impurity phase of $(Mn, Fe)_5Si_3$. The microwires possess a homogeneous distribution of chemical compositions and small grains. The maximum magnetic entropy change is 4.64 J kg$^{-1}$ K$^{-1}$ for a field change of 5 T. Relative to its bulk counterpart, the weakened magnetocaloric effect in the $Mn_{1.26}Fe_{0.60}P_{0.48}Si_{0.52}$ microwires could be associated with the weakened first-order transition. The analysis of the universal $-\Delta S_M$ ($\mu_0 H$, $T$) curve consistently reveals the presence of a weak first-order magnetic transition in these microwires.

**Acknowledgments**

This work was supported by the National Natural Science Foundation of China (NSFC, Nos. 51801044, and 51871124).

**Figure Captions**

**Fig. 1** X-ray diffraction pattern of the $Mn_{1.26}Fe_{0.60}P_{0.48}Si_{0.52}$ microwires.

**Fig. 2** (a), (b) SEM images and (c)-(g) the corresponding EDS results of the $Mn_{1.26}Fe_{0.60}P_{0.48}Si_{0.52}$ microwires.

**Fig. 3** (a) Temperature dependence of magnetization $M(T)$ taken in a field of 0.1 T for the $Mn_{1.26}Fe_{0.60}P_{0.48}Si_{0.52}$ microwires (blue experimental measuring points) and the modified Bloch's law fitting curve (the red curve). The inset shows a magnified graph over temperature range from 0 to 140 K and over magnetization range from 35 to 75 A $m^2$ $kg^{-1}$, (b) The inverse magnetic susceptibility as a function of temperature at 0.1 T (blue points) and its linear fitting curve (red curve) in the paramagnetic region for the $Mn_{1.26}Fe_{0.60}P_{0.48}Si_{0.52}$ microwires.

**Fig. 4** (a) The filled 2D contour plot of the temperature and applied magnetic field dependence of magnetization; (b) The filled 2D contour plot of the temperature and applied magnetic field dependence of $dM/dT$ for the $Mn_{1.26}Fe_{0.60}P_{0.48}Si_{0.52}$ microwires.

**Fig. 5** (a) The 3D colormap surface and (b) the filled 2D contour plots of temperature and magnetic field change (up to 5 T) dependencies of the $-\Delta S_M$ for the $Mn_{1.26}Fe_{0.60}P_{0.48}Si_{0.52}$ microwires.

**Fig. 6** The refrigerant capacity (RC) of the $Mn_{1.26}Fe_{0.60}P_{0.48}Si_{0.52}$ microwires.

**Fig. 7** (a) Arrott-Noakes plot of magnetization isotherms using the mean-field model ($\beta = 0.5$; $\gamma = 1$), (b) The universal curve $\Delta S_M/\Delta S_M^{peak}$ versus the rescaled temperatures $\theta$ of the $Mn_{1.26}Fe_{0.60}P_{0.48}Si_{0.52}$ microwires.

**Figure 1**

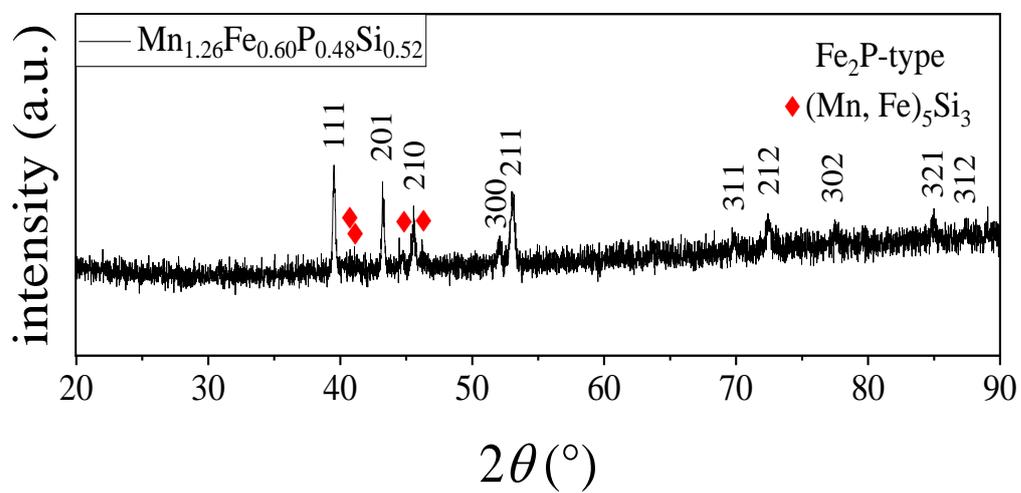

**Figure 2**

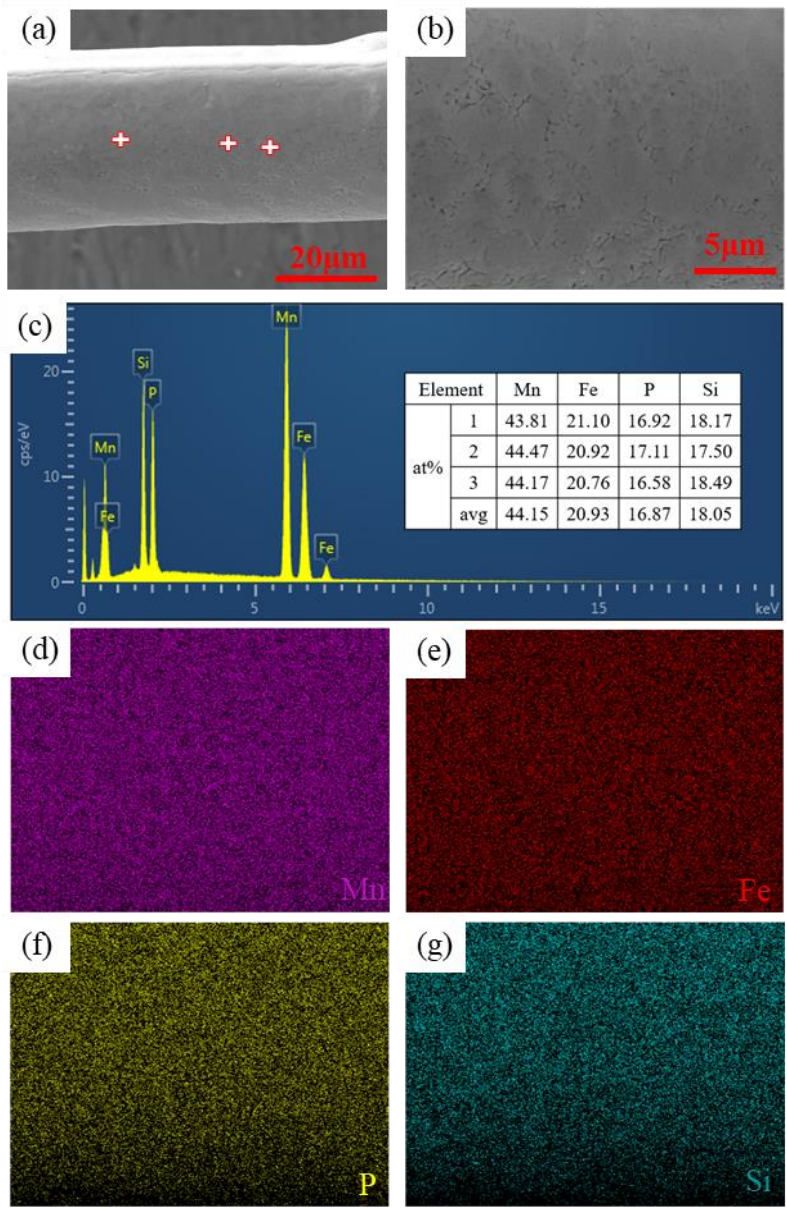

**Figure 3**

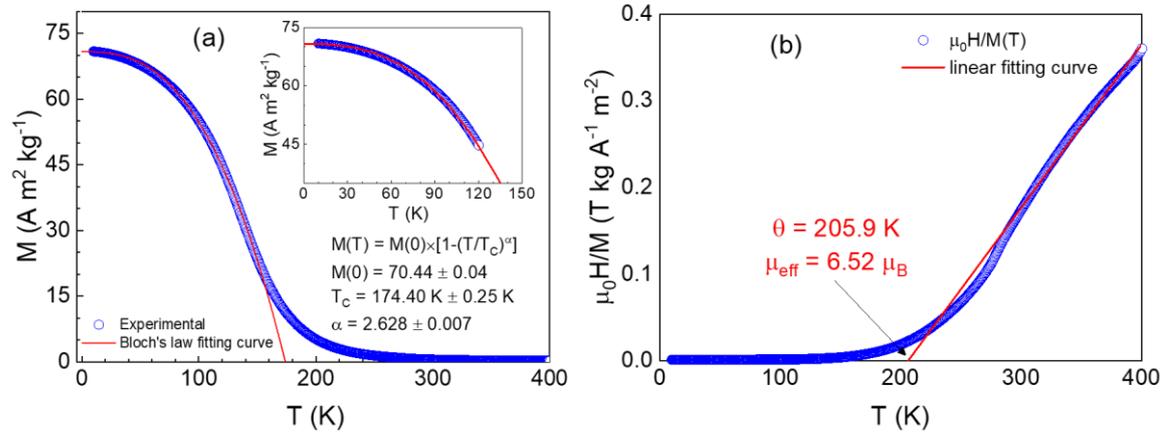

**Figure 4**

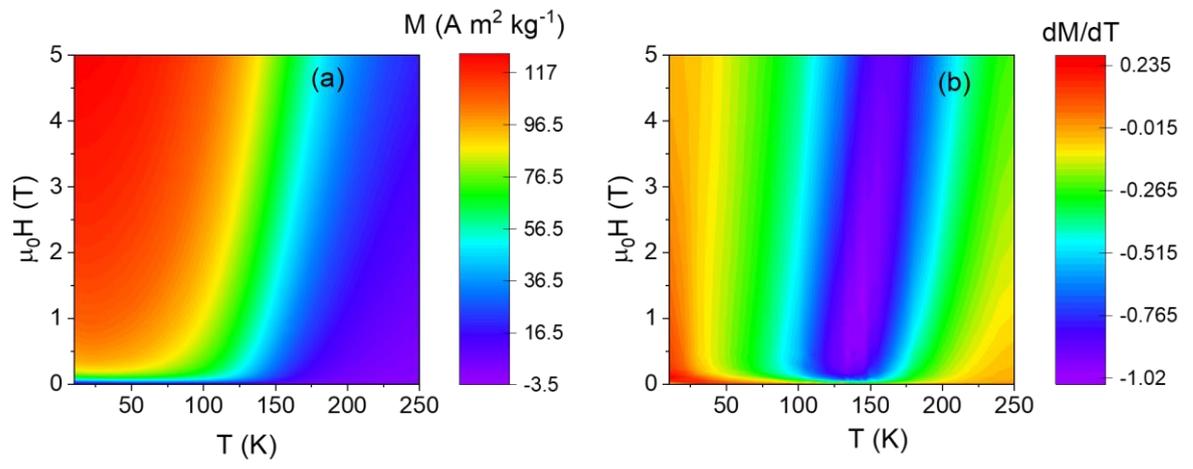

**Figure 5**

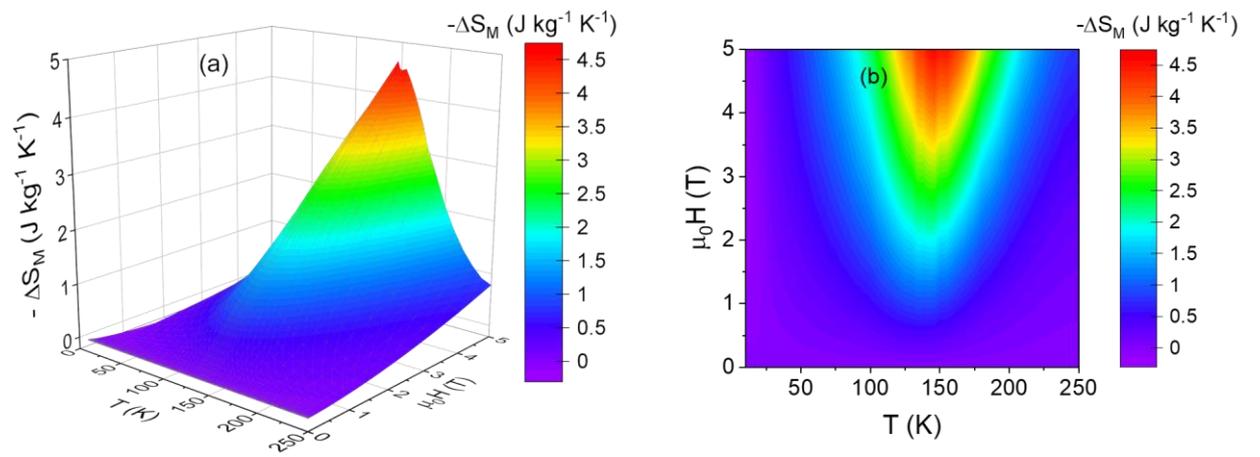

**Figure 6**

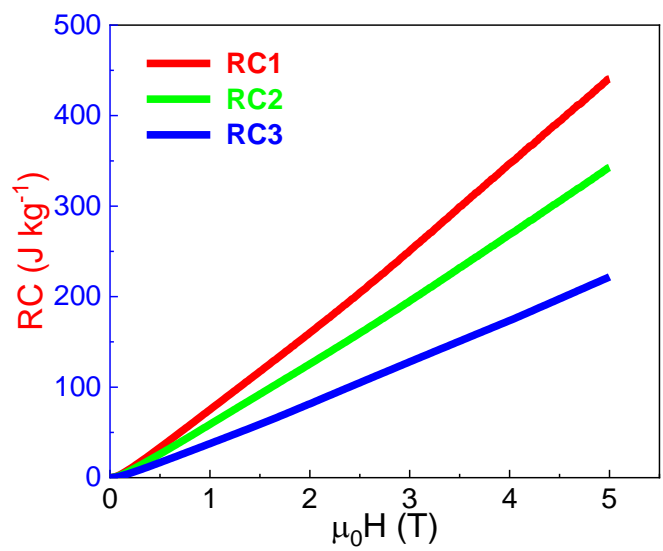

**Figure 7**

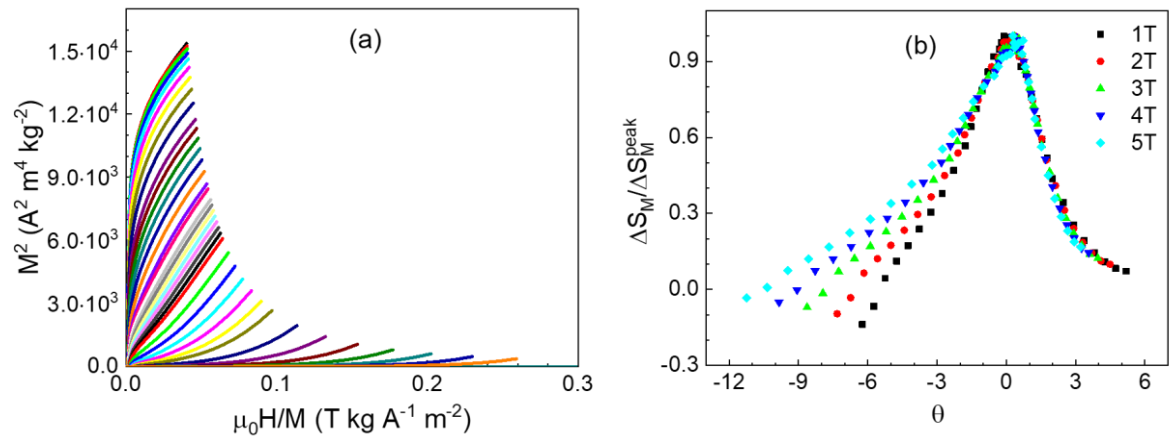